\documentclass[aip,reprint]{revtex4-1}
\usepackage{graphicx}
\usepackage{psfrag}
\usepackage{dcolumn}
\usepackage{bm}
\usepackage{latexsym}
\usepackage{amssymb}

\def\unit{\rm}
\begin{document}

\title{Feedback control of noise in spin valves by the spin-transfer torque }
\author{Swarnali Bandopadhyay}
\author{Arne Brataas}
\affiliation{Department of Physics, Norwegian University of Science and Technology,
7491, Trondheim, Norway}
\author{Gerrit E. W. Bauer}
\affiliation{Kavli Institute of NanoScience, Delft University of Technology, Lorentzweg
1, 2628 CJ Delft, The Netherlands}
\date{\today}

\begin{abstract}
The miniaturisation of magnetic read heads and random access memory elements
makes them vulnerable to thermal fluctuations. We demonstrate how current-induced 
spin-transfer torques can be used to suppress the effects of thermal fluctuations. 
This enhances the fidelity of perpendicular magnetic spin valves. The simplest
realization is a dc current to stabilize the free magnetic layers. The power
can be significantly reduced without losing fidelity by simple control
schemes, in which the stabilizing current-induced spin-transfer torque is
controlled by the instantaneous resistance. 
\end{abstract}

\maketitle


Magnetic spin valves with metal or insulator spacers are used for hard disk
read heads as well as for magnetic random access memory elements. The
physical principle of operation is the giant magnetoresistance (GMR) or
tunnel magnetoresistance (TMR), which vary with the relative magnetization
between a layer with fixed magnetization ({\it e.g.} by exchange biasing) and a
free layer with a magnetization that is allowed to (re)orient the
magnetization direction. Further miniaturization is thwarted by the
increased thermal fluctuations associated with smaller magnets, however,
because the magnetic anisotropy energies scale with the total magnetic
moment~\cite{smith01}. GMR read heads in the form of nanometer-scale pillars
are currently considered as possible replacements for TMR read heads~\cite{maat}. 
Their higher sensitivity comes at the cost of noise which is enhanced
by the spin-transfer-torque-induced coupling of electronic and magnetic
fluctuations~\cite{covingtonP,foros}. Here we demonstrate that
the spin-transfer torque can also be used to suppress the noise in spin
valves. The necessary power to achieve a given fidelity can significantly
be reduced by feedback control of the electric current-induced torque by the
instantaneous resistance.

An electric current passed through a spin valve interacts with the magnetic
order parameter when the magnetizations are non-collinear. This is caused by
the spin current component polarized normal to the magnetization of the free
layer, which is absorbed at the interface. The loss of angular momentum in
the spin current is transferred to the magnetization as a spin-transfer
torque, \cite{slon,berger,ralphR,arne00,arne01,sun} which effectively leads to magnetic
damping or anti-damping of the magnetization dynamics, depending on the
current direction. Low frequency noise can be suppressed by increasing the
magnetic anisotropy of the free magnetic layer at the cost of reduced
sensitivity. Covington~\cite{covingtonP} proposes to reduce noise by 
applying a spin-transfer torque which opposes the intrinsic damping so 
that the effective damping is reduced. According 
to the fluctuation-dissipation theorem (FDT)
the random field that perturbs the magnetization at thermal equilibrium is
indeed proportional to the Gilbert damping constant (see Eq.~(\ref{noise})
below). A reduced damping therefore reduces noise. However, in the presence of 
a spin-transfer torque driving force the
FDT does not hold and a reduced damping does not improve read-head
performance. In this Letter we prove that the opposite is true: we have to
increase the damping in order to suppress the noise. Applying a dc current
of the right direction we are able to increase the fidelity of the read head
compared to the zero current situation. The draw back of this approach is the
additional power that has to be invested for a given noise suppression. In
search of a more energy-conserving method we propose to apply a
time-dependent current that is controlled by the instantaneous magnetization
configuration as measured by the GMR/TMR. We implement here two physically
transparent feedback mechanisms, noting that performance could even be
improved by making full use of sophisticated control theory.

\begin{figure}[tbp]
\includegraphics[width=5.5cm]{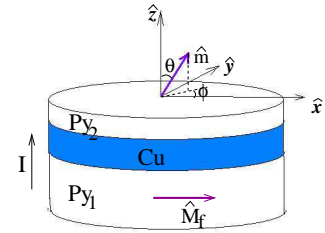}
\caption{{\protect\footnotesize (Colour online) Schematic diagram of a
perpendicular spin valve under study. A non-magnetic spacer layer (Cu) is
sandwiched between a thick ferromagnetic (Py$_{1}$) layer with `fixed'
magnetization along $\hat{M}_{\mbox f}$ and a thin ferromagnetic layer (Py$%
_{2}$) with magnetization $\hat{m}$ which is relatively `free' to move. A
positive electric charge current $\vec{I}$ is defined to flow from the
`fixed' layer to `free' layer. }}
\label{sys}
\end{figure}
Our model corresponds to a realistic~\cite{kiselev,urazhdin} nanopillar (FM$|$NM$|$%
FM) spin-valve as sketched in Fig.~\ref{sys}, consisting of a thick 
ferromagnet (FM) permalloy (Py) layer with fixed magnetization direction $\hat{M}_{%
\mbox f}$ and a thin 
Py layer separated by a non-magnetic (Cu) layer. The pillar has an
elliptical cross-section, which defines the easy axis of the free layer 
to be along $\hat{x}$. A charge current $I$ flowing from the thick layer to the thin
layer in the perpendicular direction ($\hat{z}$) is defined to be positive.
The device is stable for a dc current up to $10^{12}\,\unit{A/m^2}$ ~\cite{kiselev}. 
The allowed peak value for a pulsed current is higher~\cite{devolder08}.

We assume a spatially uniform magnetization of the free FM layer
(`macrospin' approximation). The temporal evolution of the magnetization
vector ${\hat{m}}={\vec{M}}/M_{s}$ of the free FM in presence of random
field and a Slonczewski spin-transfer torque model 
can be described by the Landau-Lifshitz-Gilbert equation: 
\begin{eqnarray}
\frac{d{\hat{m}}}{dt} &=&-\gamma {\hat{m}}\times {\vec{H}_{\mathrm{eff}}}%
+\alpha {\hat{m}}\times \frac{d{\hat{m}}}{dt}  \nonumber \\
&&-\gamma {\hat{m}}\times {\vec{h}(t)}+\gamma a_{J}{\hat{m}}\times \left( {%
\hat{m}}\times {\hat{M}_{\mathrm{f}}}\right)\,.   \label{LLG1}
\end{eqnarray}%
The effective field ${\vec{H}_{\mathrm{eff}}=-\partial }E(\vec{M})$ /${%
\partial }\vec{M}$ depends on the free energy density 
\begin{equation}
E(\vec{M})=-\frac{1}{2}\,H_{\mathrm{an}}\,M_{s}\left( \hat{m}\cdot \hat{e%
}\right) ^{2}-{\vec{H}_{\mathrm{ext}}}\cdot \vec{M}\,+2\pi M_{s}^{2}\left( 
\hat{m}\cdot \hat{z}\right) ^{2},\label{FE}
\end{equation}%
where $\hat{e}$ is the unit vector along the easy-axis.
$|{\vec{H}_{\mathrm{an}}}|=(\,N_y\,-N_x\,)\,M_s=0.05\,\unit{k\,Oe}$ is the uniaxial anisotropic field 
and $(\,N_z\,-N_y\,)\,M_s=4\,\pi\,M_{s}$ ($=5\,\unit{k\,Oe}$)~\cite{kiselev} is the demagnetization field 
(last term in the free energy Eq.~(\ref{FE})) with demagnetization factors $N_x$, $N_y$, $N_z$~\cite{osoborn} and 
$M_{s}$ is the saturation magnetization. A weak ($< 0.38\,\unit{k\,Oe}$\,\cite{kiselev}) 
external field $|{\vec{H}_{\mathrm{ext}}}| =0.1\,\unit{k\,Oe}$ is chosen to ensure no switching due to an applied 
in-plane field. The damping torque in Eq.~(\ref{LLG1}) is proportional to the Gilbert constant $\alpha =0.01$. 
The next entry on the r.h.s. of Eq.~(\ref{LLG1}) is a stochastic term. The stochastic
field ${\vec{h}(t)}$ has zero mean and white noise correlation~\cite{brown}. 
\begin{equation}
\left\langle h^{(i)}(t)h^{(j)}(t^{\prime })\right\rangle =\frac{%
2k_{B}T\alpha }{\gamma M_{s}V}\delta _{ij}\delta (t-t^{\prime })\,,
\label{noise}
\end{equation}%
where $k_{B}T$ is the thermal energy,  
$\gamma =0.176\times 10^8\,\unit{G}^{-1}\unit{s}^{-1}$ is 
the gyromagnetic ratio and $V$ the
volume of the free ferromagnet, and $\left\langle \cdots \right\rangle $
denotes statistical averaging. 
The last term on the r.h.s. of Eq.(\ref{LLG1}) is the
spin-transfer torque parameterized by $a_{J}$ which has the dimension of
magnetic field and it is proportional to applied current $I$ as
$a_{J}=(\hbar\,/2\,e\,)\,\eta\,(I/4\,\pi\,M_s\,V)$ where $\eta (=50\%)$ is the spin-polarisation.  

\begin{figure}[tbp]
\includegraphics[width=9cm]{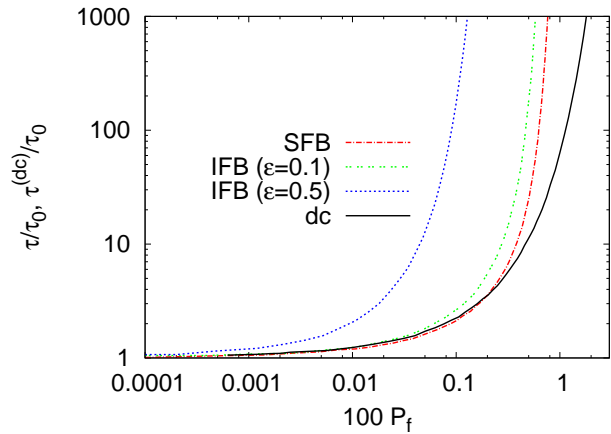} 
\caption{{\protect\footnotesize (Colour online) Fidelity life-time $\protect%
\tau $ in the presence of a dc current, as well as a pulsed current with
constant magnitude (referred as simple feedback `SFB') and varying magnitude 
(referred as improved feedback `IFB'; larger $\epsilon$ indicates stronger variation) 
in units of $\protect\tau _{0}$, the fidelity life time in the absence of a spin-torque, as a function of the
dimensionless average power consumption factor $P_f$. Results are obtained at room
temperature and averaged over $10000$ realizations. }}
\label{res1}
\end{figure}

We solve the Landau-Lifshitz-Gilbert equation numerically in the presence of 
stochastic torques at room temperature by the Heun method~\cite{garcia98}. We use discretized time 
interval $dt=1\,\unit{ps}$. We start from the initial state 
$m_{x}^{(0)} =1$ and study the temporal evolution of $m_{x}$. 
We calculate the fidelity life time, which is defined as the average time it takes
until the magnetization deviates from its initial equilibrium value beyond a
cutoff $m_{c}=0.8$ (arbitrary choice). 
The typical value of fidelity life time in absence of applied current $\tau_{0}$  
is $500\,\unit{ps}$ as obtained from our calculations. 
We average over $10000$ realizations.
 
Let us first study the dynamics in the presence of a dc current and compute the
fidelity time enhancement $\tau /\tau _{0}$ and the average
dimensionless power consumption factor $P_f=\left( a_{J}/4\pi M_{s}\right) ^{2}$ 
(the solid curve in Fig.~\ref{res1}). The dimensionless fidelity $\tau /\tau _{0}$
increases exponentially with the square of $P_f$. Thus fidelity increases with magnitude of applied current
(hence torque). As a test of our method we repeated the simulations of current induced magnetization reversal of 
Ref.~\cite{cornel06} and found good agreement. 
 
We propose two control methods to reduce power consumption,
assuming that we can monitor the instantaneous magnetization in real time by
the electrical resistance. First, we let the magnetization evolve from its
initial equilibrium configuration in the absence of a spin-torque and apply a
current pulse with constant magnitude when $m_{x}$ drops below a chosen
lower bound $m_{l}$ ($=0.85$). We switch off the current pulse
once $m_{x}$ reaches a chosen upper bound $m_{u}\,(=0.95)$. 
We refer to this method as the simple feedback scheme (SFB). 
We compute the dimensionless fidelity life
time $\tau/\tau _{0}$ and the dimensionless average power consumption factor $%
P_f=\left(\,a_{J}/4\pi M_{s}\,\right)^{2}(\Delta t/\tau)$, where $\Delta t$
is the total duration over which the spin-torque is in action. The result is shown as the dash-dotted 
line in Fig.~\ref{res1}. The fidelity life time as a function of invested power is very similar to
the dc case. In the large power regime which also refers to large current regime, the SFB leads to
significant power savings, however. A fidelity increased by a factor $1000$
compared to the zero-torque situation can be gained by investing $\sim\,3$ times
less power as compared to the dc torque method. Here our approach is to apply a larger torque 
for shorter duration as large current ensures enhancement in fidelity and small duration makes 
the scheme power-saving one. 
Note that in our numerical simulation the magnitude of current pulses would be such that the magnetization switches slowly in the time-scale of $\unit{ps}$, the chosen time-step.
In the small $P_f$ regime, the smaller enhancement in fidelity (compare
the dash-dotted curve with the solid one in Fig.~\ref{res1}) than dc case is due to the initial delay, 
set by $m_l$ on $m_x$. With increasing torque, the initial delay becomes insignificant.
       
At this stage we devise an improved control of the current pulses to save much energy. This case will be 
called as improved feedback method (IFB). As in the `SFB' case, we allow the magnetization to evolve 
from its initial state without applying a current. When $m_{x}$ drops
below $m_{l}$ we turn on a pulsed current corresponding to a torque, which now is a
function of $m_{x}$. We have chosen $a_{J}\,=a_{J}^{\left( 0\right) }\,\left[
(m_{l}-m_{c})/(m_{x}-m_{c})\right]^{\epsilon}$, where $a_{J}^{\left(
0\right) }$ corresponds to the initial value of the current pulse when it is turned on. 
$m_{l}$ and $m_{c}$ are two chosen cut-off values for $m_{x}$ as
described above. Note that in the SFB case magnitude of the pulse current is 
$a_{J}\,=a_{J}^{\left( 0\right)}$ whenever it is in action but in the present 
case $a_{J}$ increases monotonically when $m_{x}$ drops 
from $m_{l}$ and then decreases with increasing $m_{x}$ 
from $m_{c}$ in presence of spin-torque. Thus to estimate invested power we 
accumulate the power consumption factor at each time step {\it i.e.} 
$P_f=\sum_{i=1}^{N} \left(a_{J_i}/4\,\pi\,M_s\right)^2\,\left(t_i/\tau\right)$.
Since in the simulation we are using a finite time step, in practice, we do not reach the maximal current limit 
as $m_{x}\rightarrow m_{c}$. As soon as $m_{x}$ reaches $m_{u}$ we turn off the current again as in the SFB case. We carry out simulations for two values of the exponent $\epsilon$ in Fig.~\ref{res1}. For small 
$(\epsilon =0.1)$, there is only a slight improvement (the double dashed curve in 
Fig.~\ref{res1}) in the result as compared to the SFB case. For an intermediate value $(\epsilon =0.5)$, a 
significant improvement (the dashed curve in Fig.~\ref{res1}) is visible. A fidelity enhancement by a 
factor $1000$ costs roughly $15$ times less power than in the dc limit. 
Larger $\epsilon$ requires a smaller size of time-step to get reliable convergence 
in the numerical integration. 
 
We now estimate the typical power consumption of our feedback loop in a
practical situation. The average power consumption 
is $P=I^2\,R\,(\Delta t/\tau)=C_{\mbox{s}}\,P_f$ 
with $C_{\mbox{s}}=I_{\mbox{s}}^2\,R$ and $I_{\mbox{s}}=(2\,e\,V/\hbar\,\eta)\,4\pi\,M_s^2$ are system specific 
constant terms. For our system $I_{\mbox{s}}=45.5\times10^7\,\unit{e.s.u./s}\,(=0.15\unit{A})$. For a 
resistivity~\cite{counil} $\rho=70\unit{\mu\,\Omega}\,\unit{cm}$ of $2\,\unit{nm}$ thick Py at room 
temperature or resistance $R=0.75\,\unit{\Omega}$ the constant $C_{\mbox{s}}=17\,\unit{mW}$.
We consider the data from Fig.~\ref{res1} for the fidelity time $\tau =1000\tau _{0}$. For the `dc' case,
the enhancement is achieved by investing $P_f=1.8\,\times\,10^{-2}$ or average power $P=0.31\,\unit{m\,W}$. The 
same order of enhancement in fidelity is obtained by consuming $P_f=0.76\times10^{-2}$ or 
$P=0.13\,\unit{m\,W}$ in SFB scheme by applying a 
current of pulse of equivalent magnetic field $a_J/4\,\pi\,M_s=0.275$. %
For a valve with a cross-section $100\times50\,\unit{nm}^{2}$ this requires a current of magnitude 
$I=12.51\times10^7\,\unit{e.s.u./s}\,(=0.04\,\unit{A})$ which corresponds to a current density 
$8\,\times\,10^{12}\,\unit{A/m^2}$. Clearly, at such current densities our approach will break down 
due to Joule heating. 
For the IFB scheme the same enhancement in fidelity is achieved by investing 
$P_f=0.13\,\times10^{-2}$ or $P=0.02\,\unit{m\,W}$ when $\epsilon=0.5$. This strong reduction in the invested 
power will alleviate the aforementioned heating problem in SFB case. 
The efficiency (fidelity) of the operation depends
on the different system parameters (spin-torque $a_{J}/I$, external field,
anisotropy and demagnetization field etc.), but also on the chosen cut-off value 
$m_{c}$ and the boundary values $m_{l}$ and $m_{u}$.
The qualitative message appears to be valid for all sensible parameter sets,
however.

In conclusion, we propose to suppress noise in metallic magnetic spin valves
making use of the current-induced spin transfer torque. With moderate power
consumption it is possible to increase the fidelity of the
device, especially when a feedback loop is implemented which controls the
spin-transfer torque depending on the instantaneous magnetization
configuration. The latter can be measured by the electric resistance. The
correcting current pulses should have sharp rise and decay times which can
be generated by state-of-the-art electronics~\cite{ralph10}. Our method can
be easily adopted to magnetic tunneling junctions. The method can be used to
suppress noise in read heads and increase reliability of memory elements. 
\begin{acknowledgments}
This work was supported by the European Union via DYNAMAX and MACALO. SB acknowledges hospitality of TU Delft.
\end{acknowledgments}


\providecommand{\noopsort}[1]{}\providecommand{\singleletter}[1]{#1}%
\end{document}